%% file: main.tex
\documentclass[reprint,superscriptaddress,amssymb,amsmath,aps,prx,longbibliography]{revtex4-2}

\usepackage[pdftex]{graphicx}
\usepackage{xcolor}
\usepackage{mathtools,amssymb}
\usepackage{amsmath}

\newcommand{\Llen}{\mathcal{L}}
\newcommand{\Clen}{\mathcal{C}}
\newcommand{\be}{\begin{equation}}
\newcommand{\ee}{\end{equation}}

\usepackage[colorlinks=true,allcolors=blue]{hyperref}%

\begin{document}

\title{Nonlinearity and wideband parametric amplification in an NbTiN microstrip transmission line}

\author{S. Shu}
\email{shiboshu@caltech.edu}
\affiliation{California Institute of Technology, Pasadena, California 91125, USA}
\author{N. Klimovich}
\affiliation{California Institute of Technology, Pasadena, California 91125, USA}
\author{B. H. Eom}
\affiliation{Jet Propulsion Laboratory, California Institute of Technology, Pasadena, California 91109, USA}
\author{A. D. Beyer}
\affiliation{Jet Propulsion Laboratory, California Institute of Technology, Pasadena, California 91109, USA}
\author{R. Basu Thakur}
\affiliation{California Institute of Technology, Pasadena, California 91125, USA}
\author{H. G. Leduc}
\affiliation{Jet Propulsion Laboratory, California Institute of Technology, Pasadena, California 91109, USA}
\author{P. K. Day}
\email{peter.k.day@jpl.nasa.gov}
\affiliation{Jet Propulsion Laboratory, California Institute of Technology, Pasadena, California 91109, USA}

\begin{abstract}
The nonlinear response associated with the current dependence of the superconducting kinetic inductance was studied in capacitively shunted NbTiN microstrip transmission lines. It was found that the inductance per unit length of one microstrip line could be changed by up to 20\% by applying a DC current, corresponding to a single pass time delay of 0.7 ns. To investigate nonlinear dissipation, Bragg reflectors were placed on either end of a section of this type of transmission line, creating resonances over a range of frequencies.  From the change in the resonance linewidth and amplitude with DC current, the ratio of the reactive to the dissipative response of the line was found to be 788. The low dissipation makes these transmission lines suitable for a number of applications that are microwave and millimeter-wave band analogues of nonlinear optical processes. As an example, by applying a millimeter-wave pump tone, very wide band parametric amplification was observed between about 3 and 34 GHz. Use as a current variable delay line for an on-chip millimeter-wave Fourier transform spectrometer is also considered.

\end{abstract}

\maketitle

\section{Introduction}

The nonlinearity of superconducting kinetic inductance has been explored for a number of device applications including kinetic inductance travelling-wave parametric amplifiers (KI-TWPAs)~\cite{Eom:2012a}, current sensors~\cite{kher:2016a}, magnetometers~\cite{Luomahaara:2014a}, variable inductors ~\cite{Annunziata:2010a}, delay lines\cite{thakur2020superconducting}, and even qubits~\cite{Farzad:2020a}.  These nonlinear kinetic inductance devices differ in a number of ways from corresponding devices based on the nonlinear inductance of Josephson junctions.  For example, the scale of the nonlinearity for a kinetic inductor is set by the critical current, $I_c$, of a wire and is generally larger than $I_c$ for a junction.  For some types of devices, such as TWPAs, this larger current scale results in a greater dynamic range, making KI-TWPAs more suitable for high signal levels than the Josephson version of the device.  Nonlinear kinetic inductors made from large gap materials can potentially operate at higher frequency than Josephson devices based on aluminum junctions.  The maximum operating frequency is set by the gap frequency of the superconducting material, which is as high as 1.4~THz for NbTiN, compared to 90 GHz for aluminum.

A KI-TWPA was first demonstrated using a NbTiN coplanar waveguide (CPW)~\cite{Eom:2012a}, and in recent years several CPW design variations have been investigated~\cite{Vissers:2016a, Chaudhuri:2017a,Malnou:2020a}. CPWs have the advantage of a simple, single-layer fabrication process, but the necessary surrounding ground plane and the unwanted slot-line mode created by curved CPW make compact geometries difficult to realize, which impacts the achievable gain.  CPW is also less suitable for millimeter-wave applications due to radiation at bends.  In contrast, microstrip lines can be tightly meandered across a chip with little effect on performance, and superconducting microstrip lines with thin, deposited dielectrics have been used at millimeter and submillimeter wavelengths. On the other hand, the use of a deposited dielectric film potentially introduces two-level system loss~\cite{Shan:2016a,Goldstein:2020a}.  

In this paper, we present results on NbTiN microstrip transmission lines that use a hydrogenated amorphous silicon dielectric, which has shown loss tangents on the order of $10^{-5}$~\cite{OConnell:2008a, Buijtendorp:2020a,Golwala:poster}. The high dielectric constant of silicon and the use of a series of capacitive stub sections allows the characteristic impedance to be adjusted to 50 ohms, while the propagation velocity is reduced to less than $0.01 c$, allowing very compact devices.  We characterize the loss and nonlinearity in sections of this transmission line and demonstrate its use as the basis for a variable delay line for an on-chip Fourier transform spectrometer and as a very wideband parametric amplifier.

\section{Nonlinear kinetic inductance}
A supercurrent modifies the density of states of a superconductor and lowers the energy gap, resulting in an increase in the kinetic inductance~\cite{Anthore:2003a}. Due to this current dependence, the inductance per unit length of a superconducting transmission line can be expanded as

\begin{equation}
\label{eq:LI}
\Llen(I) = \Llen_0[1+(I/I_*)^2 + (I/I_*')^4 + ...]
\end{equation} 

\noindent where $I_*$ and $I_*'$ depend on the superconducting material and geometry. As the change in the phase velocity, $v_{ph} \sim 1 / \sqrt{\Llen\Clen}$, where $\Clen$ is the capacitance per unit length, varies to first order as $I^2$, this nonlinear behavior is equivalent to the Kerr effect in optical materials, in which the refractive index is intensity dependent.  When substituted into the wave equation

\begin{equation}
\label{eq:3WM}
\frac{\partial^2 I}{\partial z^2} - \frac{\partial}{\partial t}\left[\Llen(I)\Clen\frac{\partial I}{\partial t}\right] = 0 
\end{equation} 

\noindent the $I^2$ term in the expansion mixes tones at four different frequencies, which is known as four-wave mixing.  Three-wave mixing (3WM) processes may also be supported by the transmission line by applying a DC current in addition to the AC tones. In that case, $\Llen(I)$ develops a term proportional to $I$.  Either three- or four-wave mixing can produce parametric amplification.  In the case of three-wave mixing, a signal photon with frequency $\omega_s$ stimulates the conversion of a single pump photon into signal and idler photons, and the frequencies $\omega_{p,s,i}$, where the subscripts refer to pump, signal and idler tones, are related by $\omega_p = \omega_s + \omega_i$.

\section{NbTiN microstrip line}
\label{sec:msline}
\begin{figure}
    \mbox{
	\includegraphics[width=0.5\textwidth,trim={1.3cm 0cm 1cm 1cm}]{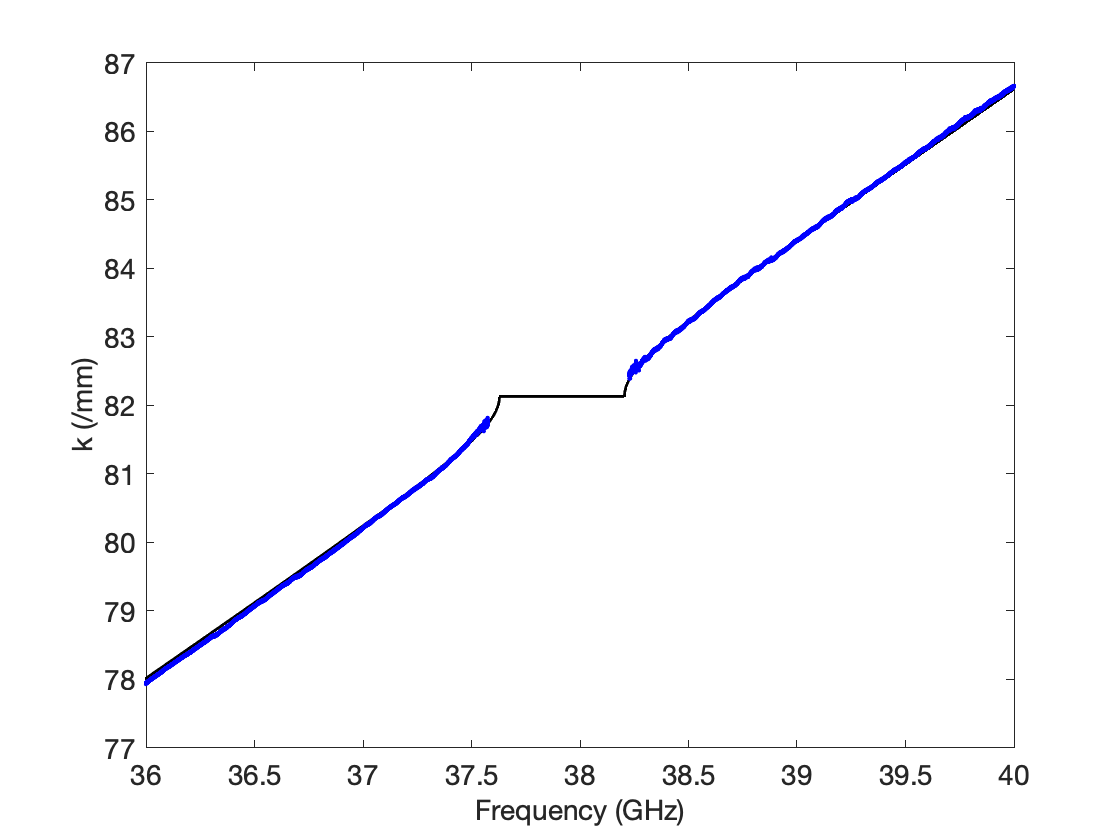}
	\hspace{-8cm}
	\raisebox{4.5cm}{\includegraphics[width=4.2cm]{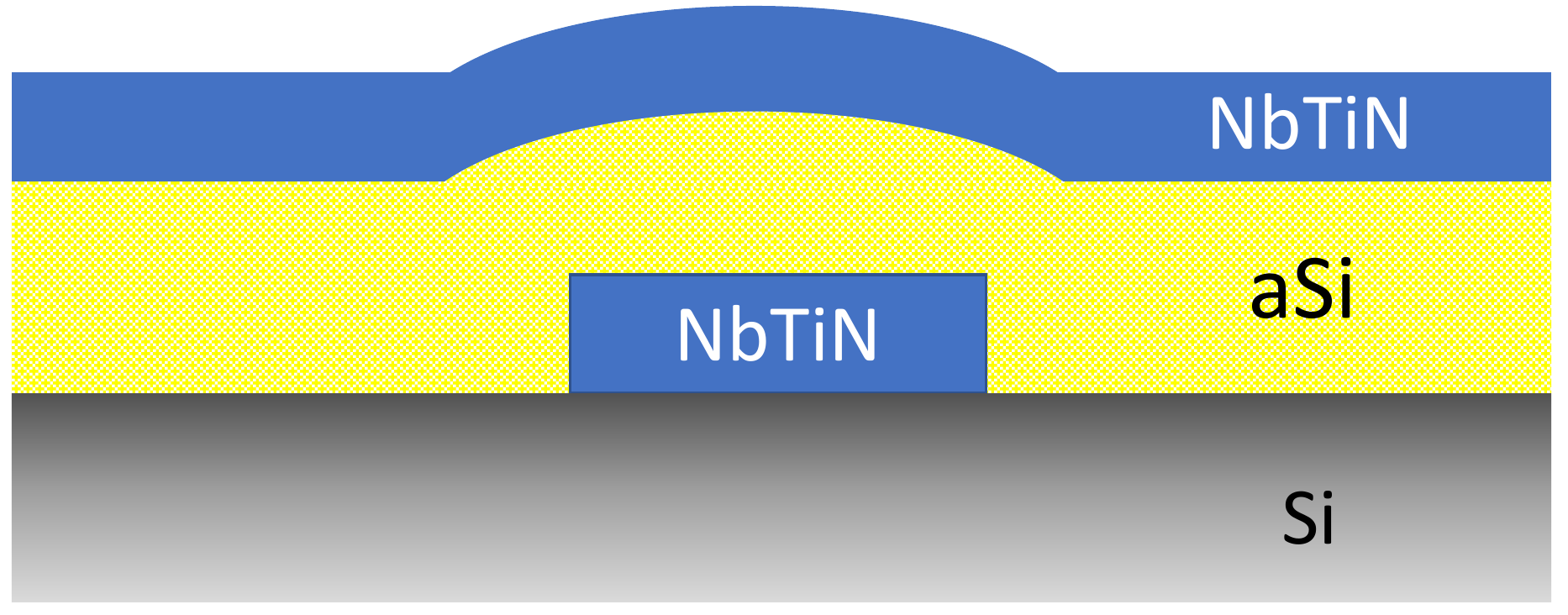}}
	\hspace{-1.95cm}
	\raisebox{8mm}{\includegraphics[width=4.6cm]{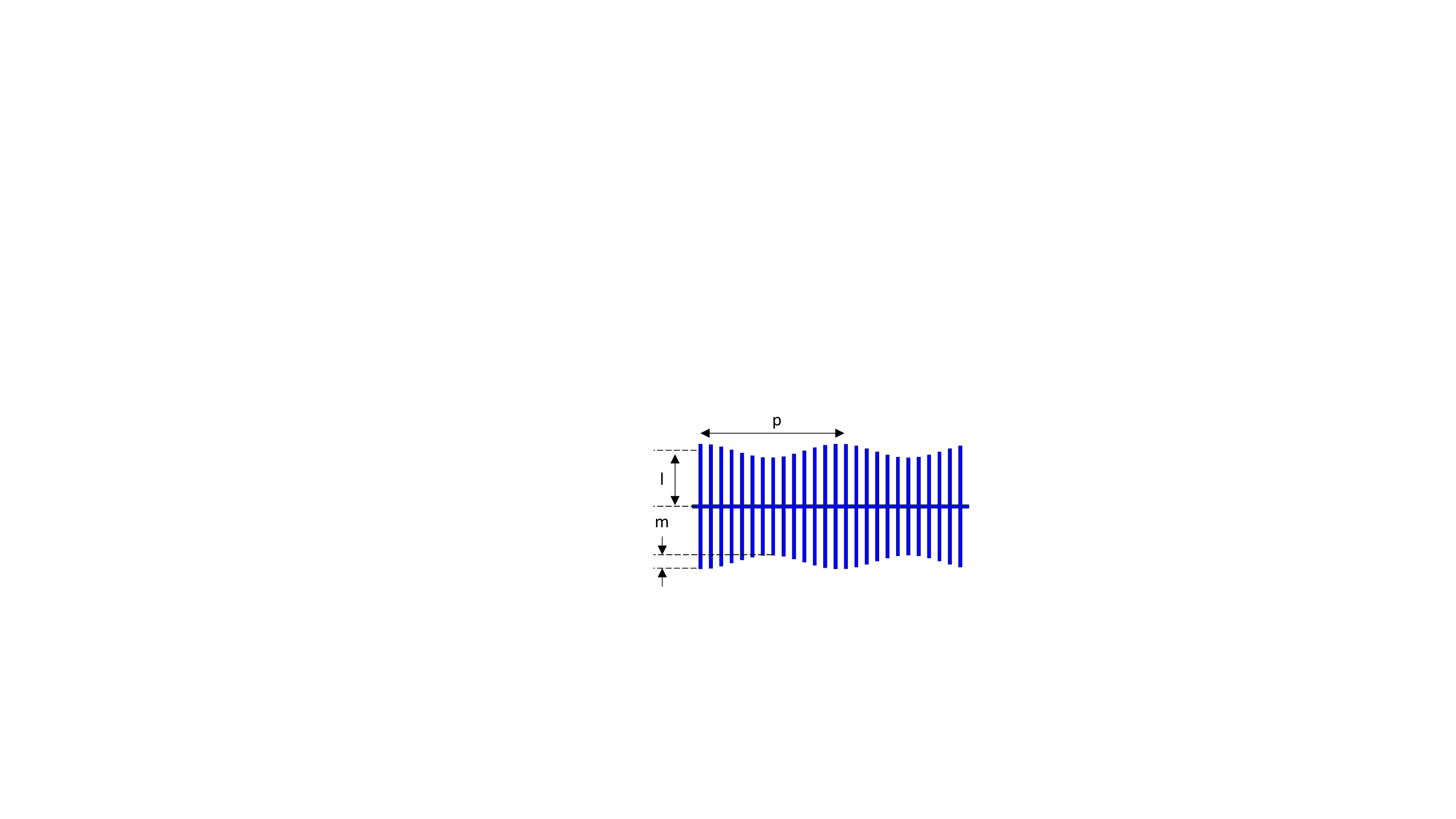}}}
	\caption{The dispersion curve around the band gap frequency of the stub-loaded NbTiN microstrip line with modulation length $p=38\mu$m, mean stub length $l=3\mu$m and stub length modulation $m=0.2\mu$m.  The measurement (blue) was taken with a DC bias current of 0.57~mA and is compared with the result of a circuit model (black), where the penetration depth has been adjusted to match the 38~GHz bandgap feature created by the periodic modulation. The upper inset shows the layers that form the transmission line.  The lower inset shows the arrangement of the microstrip stubs.  The width of all of the lines is 0.25$\mu$m.}
	\label{fig:BandGap}
\end{figure}

The geometry of the microstrip transmission lines is described in the insets of Fig.~\ref{fig:BandGap}.  Both versions of the transmission lines that were studied use a 35~nm NbTiN conductor layer that is deposited first on a high resistivity silicon substrate and patterned to form the microstrip conductor.  An amorphous silicon layer is deposited next and serves as the microstrip dielectric with dielectric constant of 10.3.  A 200~nm NbTiN layer deposited last forms for the ground plane of the inverted microstrip structure.  To maximize the nonlinear response (minimize $I_*$), a relatively narrow conductor line width of 250~nm is used. In order to maintain a characteristic impedance of 50$\Omega$ for convenient matching to external circuitry, a series of microstrip open stubs is connected to the transmission line.  At the operating frequency, the stubs are much shorter than a quarter wavelength and approximate shunt capacitances.  As a result, the final microstrip line has both a large capacitance and inductance per unit length, which reduces the phase velocity so that $v_{ph} \sim 1/\sqrt{\Llen\Clen} < 0.01c$.  The slow-wave nature of the transmission line effectively increases the electrical length and results in a compact device. For example, this design can shorten the device length from 2~m in Ref.~\cite{Vissers:2016a} to 0.1~m by a factor of 20.

An estimate of the nonlinear current scale $I_*$ can be made by equating the kinetic energy $\Llen_{kin} I_*^2/2$, where $\Llen_{kin}$ is the kinetic inductance per unit length of the microstrip conductor, with the condensation energy $N_0 \Delta^2 wt/2$, where $N_0$ is the density of states at the Fermi level and $w$ and $t$ are the width and thickness of the conductor.  For $t \ll \lambda_L$, where $\lambda_L$ is the London penetration depth, and $w \ll \lambda^2 / t$, the current is nearly uniform throughout the conductor cross section, and $\Llen_{kin} = \mu_0 \lambda_L^2 / wt$.  The nonlinear current scale can then be expressed as
\begin{equation}
    I_* = wt\kappa_*\sqrt{\frac{N_0\Delta^2}{\mu_0 \lambda_L^2}},
\end{equation}
where $\kappa_*$ is of order one.  An more rigorous analysis based on the Usadel theory\cite{Anthore:2003a,Zhao:2020a} suggests $\kappa_* = 1.37$.

$I_*$ is reduced for materials with high normal state resistivity which implies large $\lambda_L$.  The NbTiN films used for this study were produced by reactive sputtering and have resistivity in the range of 200 $\mu\Omega$cm. For a film thickness of 35~nm, we obtain a surface inductance $L_s \approx 7$~pH.  The relatively high  critical temperature, $T_c \sim 12.5$~K for a 35~nm film, allows for convenient testing with minimal loss at 1~K. 

In a traveling wave structure, the efficiency of a nonlinear process, such as parametric amplification, is controlled by the degree to which the waves involved maintain a specific phase relation. The frequency range over which phase matching can be achieved determines the bandwidth of the process.  The dispersion of the transmission medium is a main factor in phase matching. Superconducting transmission lines operated well below the gap frequency have very little intrinsic material dispersion, which can result in several nonlinear processes being simultaneously phase matched. For a particular application, the phases of waves corresponding to unwanted nonlinear processes, eg. harmonic generation in the case of parametric amplification, should ideally be mismatched so that those processes are suppressed.  Phase matching can be controlled by adding geometrical dispersion to the transmission line structure, a process referred to as ``dispersion engineering''.  The open stubs of the transmission line structure shown in Fig.~\ref{fig:BandGap} have a dispersive effect because they add a frequency dependent admittance per unit length proportional to $\tan (\omega l / v_{ph})$, where $l$ is the finger length and $v_{ph}$ is the propagation velocity on the microstrip line forming the stub.  The dispersion of the transmission line may be further modified by adding a periodic modulation, as was first implemented for a TWPA in ref \cite{Eom:2012a}. Similar approach has been applied in Josephson-junction array based TWPAs~\cite{Planat:2020JPAbuisson}. As shown in the Fig.~\ref{fig:BandGap} inset, the length of the stubs is sinusoidally varied with period $p$ producing a bandgap at frequency $v / 2p$, where $v$ is the propagation velocity on the stub loaded microstrip structure with average stub length $l$. The dispersion is modified near the bandgap frequency. The amplitude of the sine-wave  modulation sets the frequency width of the bandgap and also the degree to which the dispersion deviates near the bandgap edges.

\section{Loss measurement using an on-chip interferometer}
\label{sec:loss}
\subsection{On-chip interferometer design}
\begin{figure}
    \centering
    \includegraphics[width=0.45\textwidth]{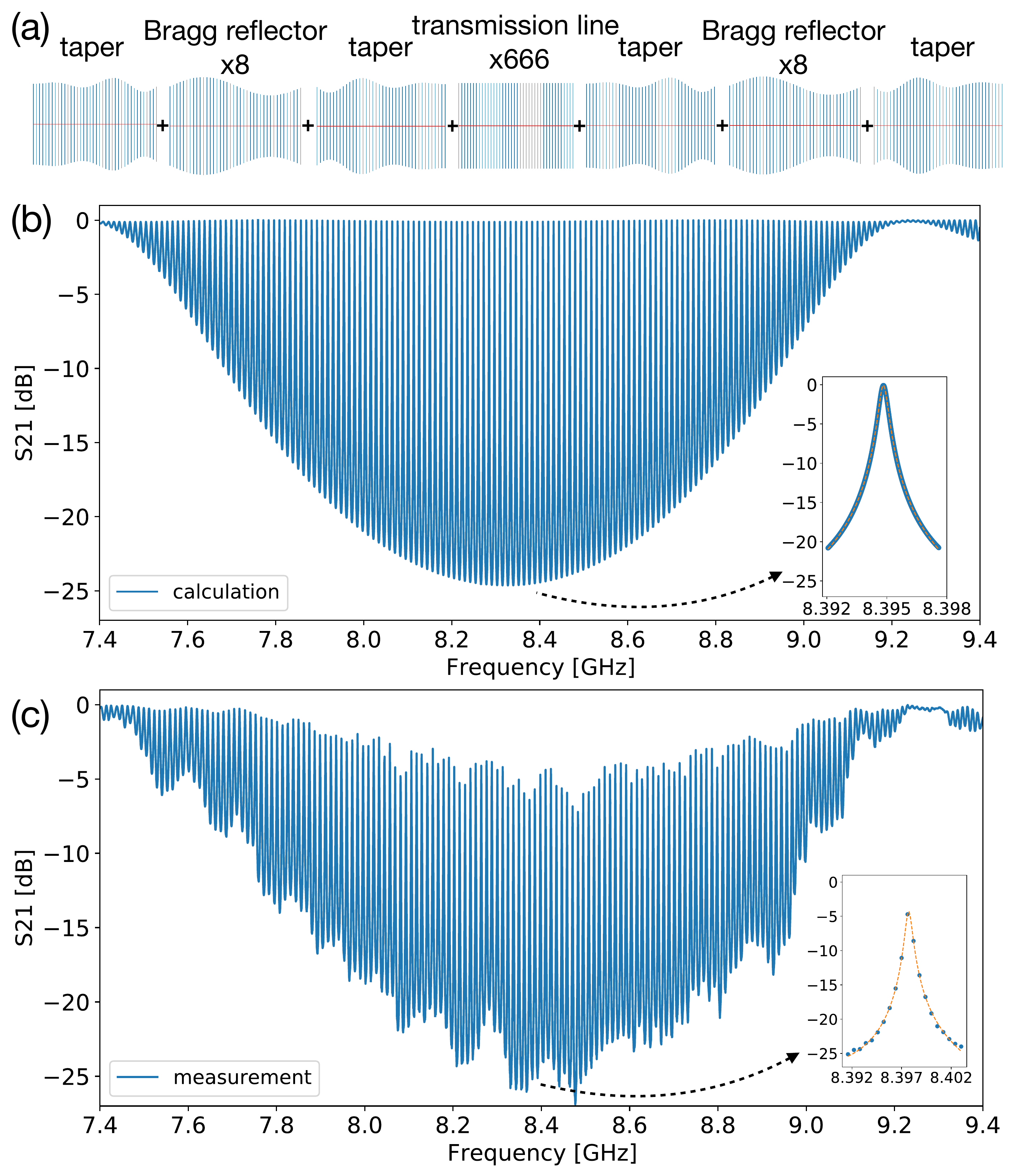}
    \caption{(a) The schematic structure of the on-chip interferometer device. (b) The calculated S21 of the device without loss. (c) The measured S21 at $1$ K after baseline correction. The insets show the Lorentzian fit of resonators.}
    \label{fig:Loss_S21}
\end{figure}

To measure the loss in the transmission line, we designed an on-chip Fabry–P\'{e}rot interferometer at $8.3$ GHz. The device consists of a stub-loaded transmission line of length $L=93$~mm and two Bragg reflectors, located at the two ends of the transmission line, as shown in Fig.~\ref{fig:Loss_S21}. The stub length in the reflector sections has a sine wave variation starting from $l=26$ $\mu m$ with an amplitude of $m=12$ $\mu m$ and a period of $p=140$ $\mu m$, and the transmission line between the reflectors has a uniform stub length of $l=26$ $\mu m$.  The transmission line is arranged in a meandering pattern across a 2.5~mm $\times$ 26~mm chip . The spacing between the stubs is $2$ $\mu m$ for the whole device, and the aSi dielectric thickness is 190~nm. The reflector has a length of $8$ periods of the stub length modulation, which produces an incomplete stop band around $8.3$ GHz. To decrease the impedance mismatch between the reflectors and the transmission line near the stop band, there sections of $2$ periods of length over which the modulation amplitude is tapered in and out at the input and output of each reflector. The device acts as an etalon for frequencies within the stop band, resulting in transmission peaks at frequencies $\omega_n = n \pi v / L$. From the measured frequency spacing (Fig.~\ref{fig:Loss_S21}), we find that $v=0.0077c$ for this device for $I_{DC}=0$.  The transmission of the etalon may be expressed as
\begin{equation}
    \mathrm{S21} = \frac{t^2 e^{-\gamma L}}{1 - r^2 e^{-2\gamma L}},
\end{equation}
where $t$ and $r$ are the frequency dependent Bragg reflector transmission and reflection amplitudes, $\gamma = \alpha + i\beta$ is the propagation constant on the internal transmission line section, and $\beta = \omega / v$.  Near the resonance frequencies, the transmission peaks are nearly Lorentzian and
\begin{equation}
    \mathrm{S21}(\omega_n + \Delta\omega) \approx \frac{Q_r}{Q_c}\frac{t^2}{1-r^2} \frac{ e^{-\gamma L} }{1 - 2iQ_r\Delta\omega / \omega_n},
\end{equation}
where $Q_r$ is the full width at half maximum of the resonance and $Q_c = (\omega_n L/v) r^2 / (1 - r^2)$ is a measure of losses to the external circuit.  Internal losses are represented by $Q_i = \beta / 2\alpha$, and the quality factors are related by 
\begin{equation}
    Q_r^{-1} = Q_c^{-1} + Q_i^{-1}.
    \label{eq:Qs}
\end{equation}  
Neglecting loss in the Bragg reflectors ($r^2 + t^2 = 1)$ and a phase factor related to the phases of $r$ and $t$, and normalizing the transmission to that of a single pass through the internal transmission line section, the transmission becomes
\begin{equation}
        \mathrm{S21} \approx \frac{Q_r}{Q_c} \frac{ 1 }{1 - 2iQ_r\Delta\omega / \omega_n}.
        \label{eqn:res}
\end{equation}

\subsection{Loss measurement}
The microwave transmission of the device was measured at $1$ K with a series of DC bias currents. The calculated and measured S21 with $I_{DC}=0$ are shown in Fig.~\ref{fig:Loss_S21}. The calculation was made by cascading the ABCD matrices of the microstrip sections making up the device.  The measured S21 is normalized by subtracting a linear-in-dB baseline found by fitting S21 at frequencies above and below the stop band.  

\begin{figure}
    \centering
    \includegraphics[width=0.45\textwidth,trim={0cm 0cm 0cm 0cm},clip]{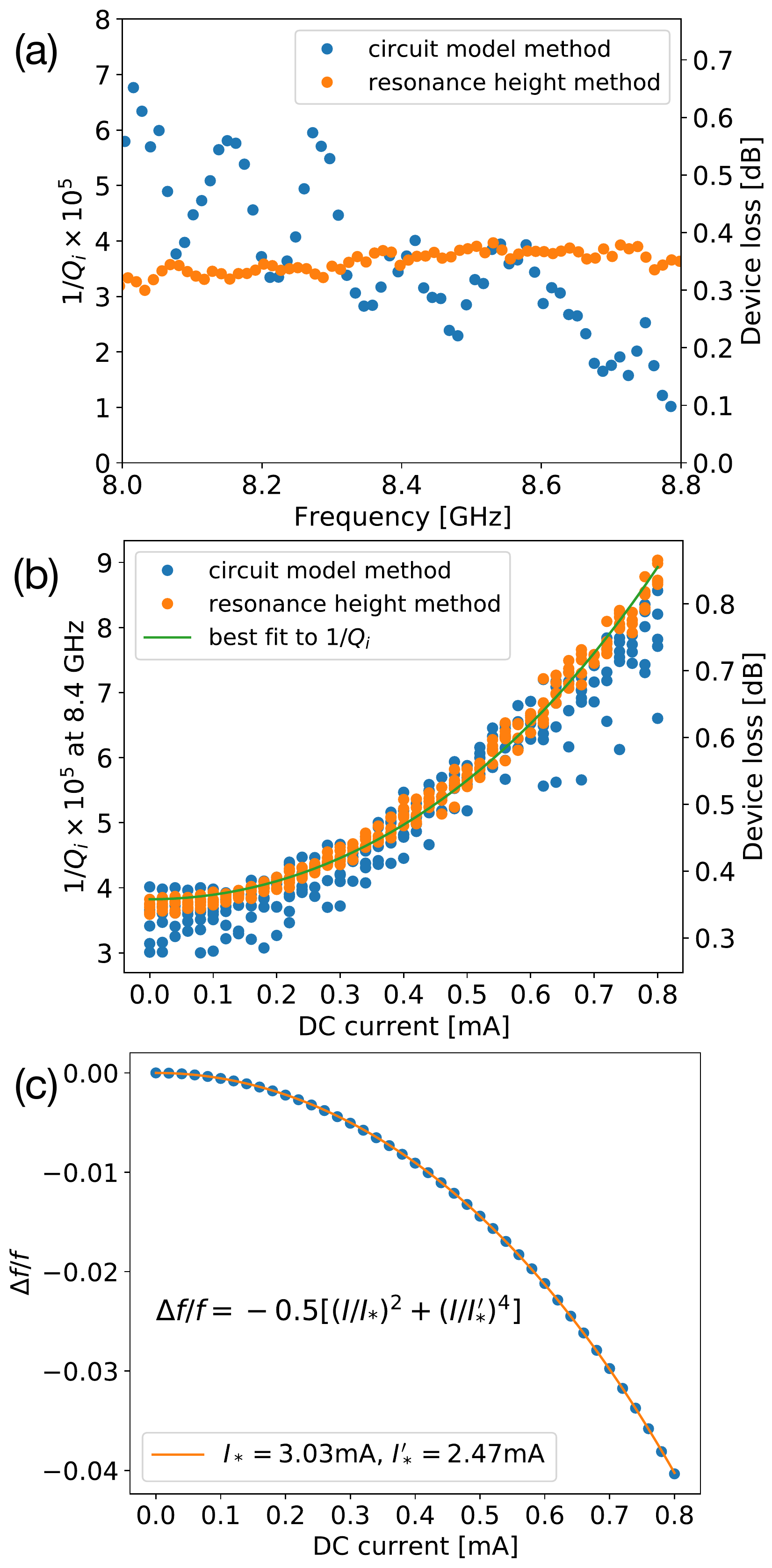}
    \caption{(a) The frequency dependent attenuation factor without DC bias. (b) The bias current dependent attenuation factor. The device loss is calculated assuming a one-way parametric amplifier using this transmission line with the same length, $93$ mm. (c) The frequency shift of a single resonance by increasing DC current. The fitted scale of the nonlinearity is $I_*=$ 3~mA.}
    \label{fig:Loss_att}
\end{figure}

Resonances were fit to a Lorentzian function to extract the quality factors $Q_r$ in both calculation and measurement. Losses were not included in the calculation, so in that case $Q_r = Q_c$.  To find the experimental $Q_i$ from the measured $Q_r$, two methods are used. In the ``circuit model method'',  the $Q_c$ derived from the S21 calculation is used along with Eqn.~\ref{eq:Qs}. The second method uses the resonance height in the measured, baseline-normalized S21 to calculate $Q_i$. According to Eqns.~\ref{eq:Qs} and \ref{eqn:res}
\begin{equation}
    Q_i = \frac{Q_r}{1-\left | \mathrm{S21}_{\mathrm{max}}\right |},
\end{equation}
where $|\mathrm{S21}_{\mathrm{max}}|$ is the maximum $|\mathrm{S21}|$ of each resonance. Because the $Q_c$ are higher in the center of the stop band, the central resonances are more sensitive to $Q_i$. The attenuation factors and the total one-pass loss of the intermediate transmission line section are plotted in Fig.~\ref{fig:Loss_att} using the data from resonances near the stop band center.

The attenuation calculated using the circuit model method has a large variation and decreases as frequency increases, suggesting some discrepancy between the actual and designed $Q_c$ values. The loss found with the baseline method has less variation and increases slightly with frequency. An increase in loss with frequency is expected from the increase in the phase length.  Around $8.4$~GHz the two methods give consistent results, $\alpha=3.7$~dB/m with $I_{DC}=0$. The one-pass loss of this $93$~mm transmission line, which corresponds to 333 wavelengths, is $0.35$ dB and $Q_i=2.8\times10^4$.  The loss is higher than was found in resonator measurements using the same dielectric material where $Q_i > 10^5$ is observed and resonance frequency shift versus temperature measurements imply a single photon loss factor of $10^{-5}$\cite{Golwala:poster}. It is possible that the extra loss comes from fabrication defects along the length of the device, which is much longer than the resonators that were studied. 

The transmission was also measured as a function of DC current injected through the length of the transmission line up to $0.8$~mA. Past that current, the line became resistive.  The same two methods are used for calculating the attenuation factors, shown in Fig.~\ref{fig:Loss_att}. The stop band and resonance frequencies shift to lower frequency when the current is increased, so the resonance closest to 8.4~GHz at each measurement current was used to avoid mixing in the frequency dependence of the loss.  Both methods give consistent results at $I_{DC}<0.6$~mA. When $I_{DC}>0.6$~mA, the circuit model method gives lower values, most likely because the 8.4~GHz resonance moves to the edge of the stop band, where that method produced less stable results.

The $Q_i$ decreases by a factor of 2.3 from $2.8\times 10^4$ at zero DC current to $1.2\times 10^4$ with $I_{DC}=0.8$~mA, while the device loss increases by a factor of 3.4 from 0.35~dB to 0.84~dB. The device loss increase includes a contribution from the increase in the electrical length of the transmission line. We define the ratio of the reactive to the dissipative response, $R=\Delta\beta/2\Delta\alpha$. Therefore, we have
\begin{equation}
    \frac{1}{Q_i}=\frac{2\alpha}{\beta} = \frac{1}{Q_i(0)} + \frac{1}{R} \frac{\Delta\beta}{\beta},
    \label{eqn:lossratio}
\end{equation}
where $\beta=\omega/v$ and $v = 0.0077c[1-0.5(I/I_*)^2-0.5(I/I'_*)^4]$. The values of $I_*$ and $I'_*$ are obtained from fitting $\Delta f/f$ (Fig.~\ref{fig:Loss_att}). Fitting the resonance height method loss data to Eqn.~\ref{eqn:lossratio} yields $R=788$.

The origin of the decrease of $Q_i$ with DC current is not clear.  Using the Usadel equations and Nam's equations~\cite{Usadel:1970a,Nam:1967a,Zhao:2020a,Driessen:2012a}, we calculated the conductivity quality factor $Q_i=\sigma_2/\sigma_1>10^{8}$ at $I_{DC}=0.8$~mA, so the measured $Q_i$ is not limited by the current induced change of the density of states. The $Q_i$ is also relatively insensitive to temperature at 1~K, so microwave heating of the transmission line should have negligible effect. It is possible that the loss happens locally due to the fabrication nonuniformity. In a separate measurement, the degradation of $Q_i$ is also negligible under parallel magnetic field up to 40~mT by putting the device inside a solenoid. It has been shown that the $Q_i$ of a 250~nm width NbTiN resonator decreases from $10^5$ to $10^4$ with perpendicular magnetic field from 0 to 150~mT~\cite{Samkharadze:2016NbTiNMag}. In our measurement, no external magnetic field was applied and the magnetic field generated by the DC current is 0.84~mT, suggesting loss in the microstrip conductor strip is negligible. However, the NbTiN ground plane covers the whole chip and the electromagnetic field extends to a wider area than the strip width. The degradation of $Q_i$ from $10^5$ to $10^3$ with a perpendicular magnetic field up to 3~mT has been reported~\cite{Kroll:2019NbTiNMag}, so it is also possible that our $Q_i$ is limited by itinerant vortices and the increased current creates more vortices in the ground plane. As the magnetic field is quite small, the related $\Delta f/f$ is at the level of $10^{-3}$, which is small compared with the current induced one.

As a feature of phase-sensitive amplifiers, parametric amplifiers are able to squeeze the quantum fluctuation\cite{Castellanos:2008squeeze}. Therefore, we further investigate how the asymmetric loss impacts the squeezing performance of a KI-TWPA in 3WM regime. Using the measured $Q_i=2.8\times10^4$ and assuming $Q_i$ is constant at all frequencies, a small amount of loss asymmetry still exists in non-degenerate amplification, due to the different electrical lengths for signal and idler frequencies. Assuming signal at 6~GHz and idler at 8~GHz, the loss gives a maximum squeezing level of 25~dB. The loss asymmetry sets the optimal length of the amplifier to be $\sim 50$~mm. The calculation details are discussed in Appendix~\ref{app:asym}.

\section{A multi-band parametric amplifier}

\subsection{Device design}
\begin{figure}
    \centering
    \includegraphics[width=0.45\textwidth]{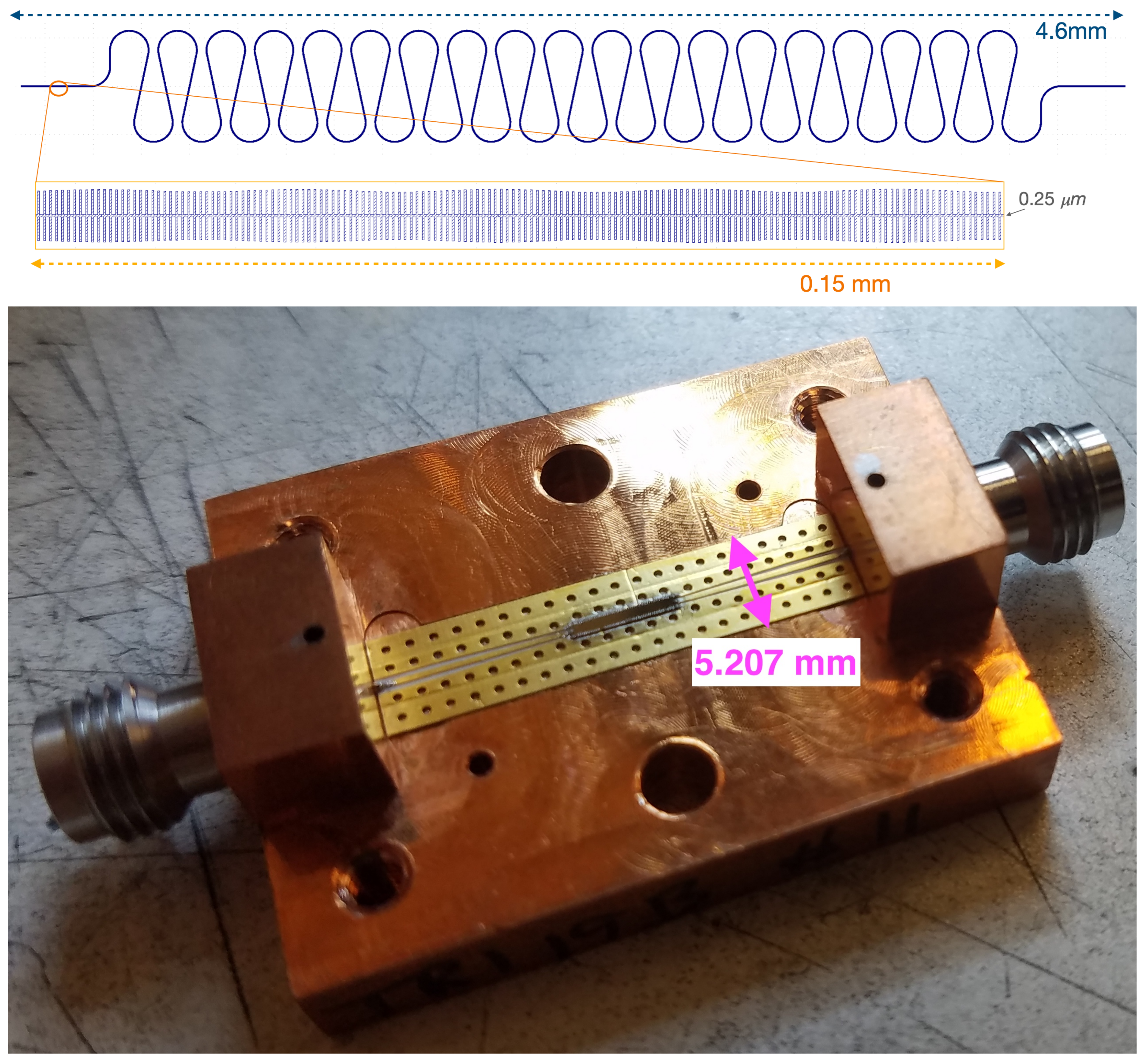}
    \caption{The upper panel shows the layout of the multi-band KI-TWPA, which is arranged in a meandering path with a total length of 21~mm. The zoom-in shows the modulation of capacitor fingers, which introduce the frequency bandgap. The chip is 0.6 mm $\times$ 4.6 mm and sits on a circuit board in a holder.  A lid (not shown) forms a channel around the chip with a cut-off frequency above the operating band. }
    \label{fig:paramp_photo}
\end{figure}

A second stub-loaded microstrip line that was studied was designed for use as a wideband KI-TWPA. The device has a similar layer setup as the microstrip line in Section~\ref{sec:msline}, except that the amorphous silicon dielectric layer thickness is decreased to 60~nm and the substrate thickness is decreased to 100~$\mu$m. The NbTiN thicknesses were nominally the same as the earlier device, but a different sputtering target was used and the deposition rate was different. The stubs have an average length of $l=3~\mu$m, a modulation amplitude of $m=0.2~\mu$m, and a modulation period of $p=38.25~\mu$m, producing a bandgap at 38.5~GHz, as shown in Fig.~\ref{fig:Ka_rawS21}. The transmission line is arranged in a meandering pattern with a total length of $L=21$~mm across a $4.6\times 0.6$~mm chip, shown in Fig.~\ref{fig:paramp_photo}.

With this compact design, this device can be further integrated with other on-chip circuits to make a multi-pixel design for astronomical observations.

\subsection{Gain Measurements}
\label{sec:gain}
The gain measurements were made at 1~K. A DC current was injected through the transmission line along with a microwave pump tone to permit parametric amplification via 3-wave mixing (3WM). This process results in the formation of an idler tone, with pump, signal and idler frequencies related by $\omega_i = \omega_p - \omega_s$.  The gain of the amplifier is maximized when the dispersion of the transmission line is tuned to provide a phase mismatch, characterized by $\Delta k = \beta_p - \beta_s - \beta_i$, with wavevectors $\beta_n = \beta(\omega_n$), between the tones that compensates the nonlinear phase difference due to self and cross-phase modulation\cite{Malnou:2020a}:
\begin{equation}
\label{eq:phasematch}
    \Delta k = -\frac{ I_{p}^{2}}{8I_\dagger^2}\left(\beta_{p}-2 \beta_{s}-2 \beta_{i}\right).
\end{equation}
where $I_\dagger^2 = I_*^2 + I_{DC}^2$. 

\begin{figure}
	\centering
	\includegraphics[width=0.45\textwidth,trim={0cm 0cm 0cm 0cm}]{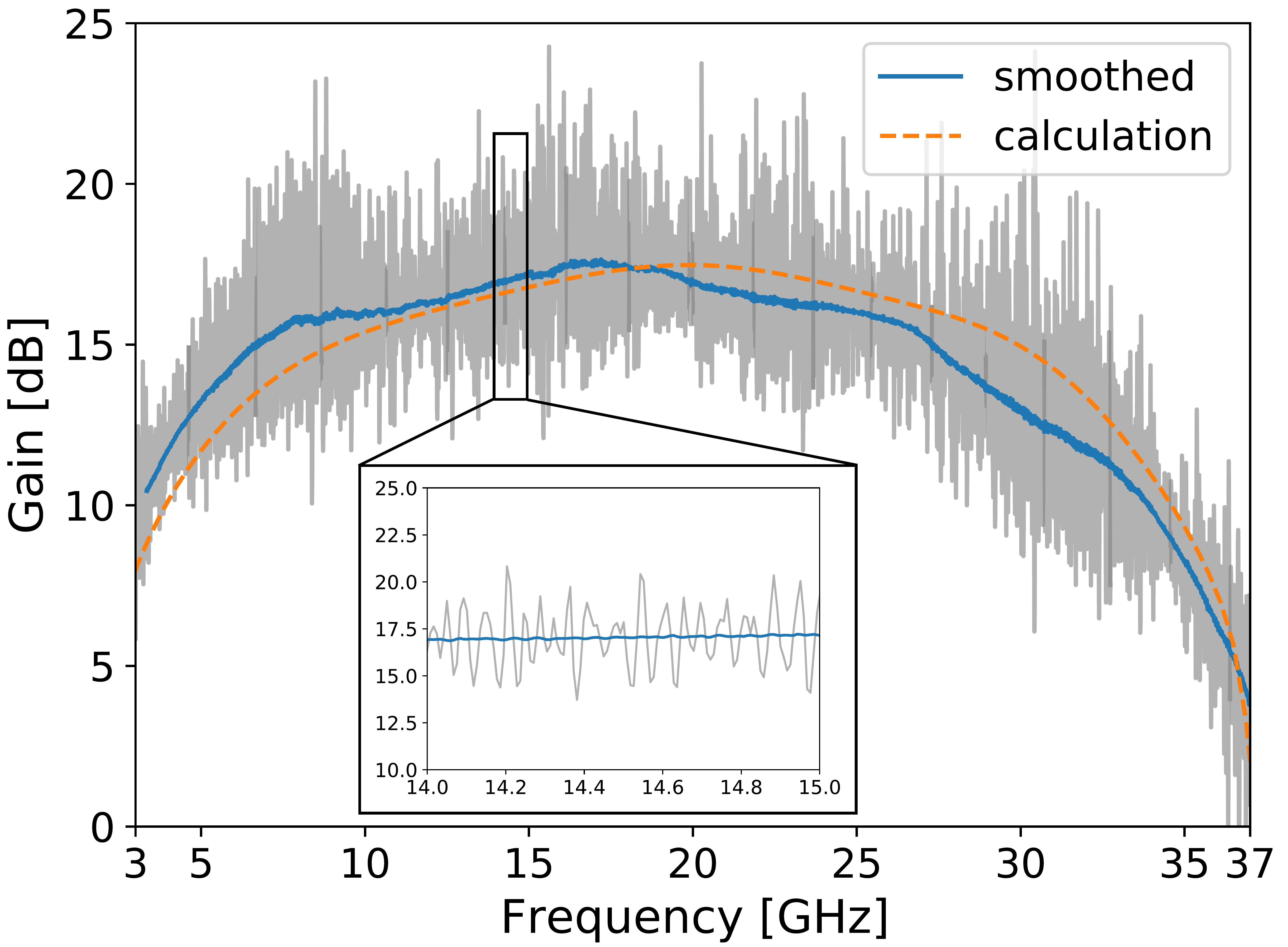}
	\caption{Gain of the parametric amplifier gain at 1~K with a -29.3~dBm pump tone at 38.8~GHz and $I_{DC} = 0.75$~mA. The gray line is the measured result and shows a large ripple.  The blue curve is the data averaged in dB over frequency.  The orange curve is the result of the couple-mode calculation.  The inset shows ripple over a narrow frequency range.}
	\label{fig:gain}
\end{figure}

The gain of the amplifier with a DC bias and pump tone applied is shown in Fig.~\ref{fig:gain}.  Here the gain is defined as the ratio of the output power with the pump on to that with the pump off.  Based on the results of section \ref{sec:loss}, a small frequency dependent loss of approximately 0.01~dB/GHz is expected through the device, which is not included in the reported gain.  The device produces gain over a wide bandwidth from 3 to 34~GHz with a pump frequency of 38.8~GHz, which lies just above the bandgap at the bias current $I_{DC} = 0.75$~mA.  A calculated gain curve derived from integrating coupled mode equations is shown in Fig.~\ref{fig:gain} for comparison with the measured result.  We found that including nonlinear processes in the calculation in addition to the 3WM gain process improved the match to the experimental result.  The other processes included were 2nd and 3rd harmonic generation and four-wave mixing (4WM) parametric amplification.  These additional processes can be expected to occur because the tones at the frequencies involved are not strongly mismatched.  The coupled mode equations used for the calculated gain profile are explained in Appendix \ref{sec:coupledmode}.  This method uses the
dispersion curve for the stub-shunted, modulated transmission line, calculated using the circuit model of the device, as an input.  The difference between the calculation and measurement suggests that some parameters may deviate from our model. The 1~dB compression occurs above -57~dBm with 18~dB gain, measured using a similar device.

\section{Mm-wave current variable delay line}
\subsection{Current dependent transmission}
\begin{figure}[h!]
    \centering
    \includegraphics[width=0.47\textwidth]{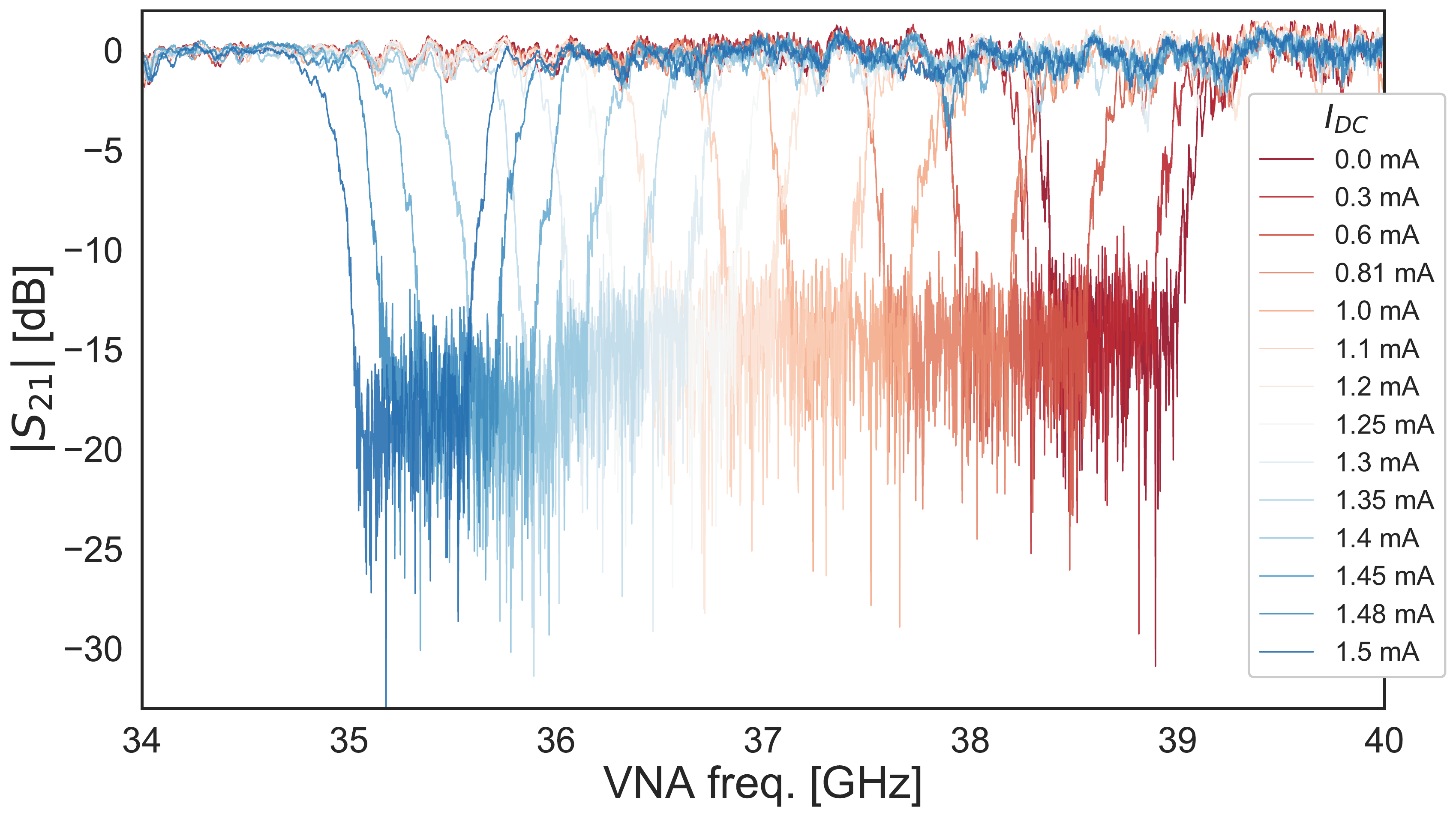}
    \caption{S21 of the multi-band KI-TWPA measured at 1~K at different DC current bias.  The transmission is normalized by subtracting a linear baseline determined by fitting the data outside the bandgap.}
    \label{fig:Ka_rawS21}
\end{figure}

In this section, we characterize the same device without the pump signal for use as a millimeter-wave current variable phase delay element. With only a DC current applied to the transmission line there is no amplification, but the propagation velocity changes as a result of the current induced change in kinetic inductance.  We measure the propagation velocity and inductance change by tracking the frequency of the bandgap that is created by the 38~$\mu$m stub length modulation.  The millimeter-wave transmission through the device near the bandgap frequency is shown in figure Fig.~\ref{fig:Ka_rawS21} for a series of DC currents. 

The bandgap frequencies $\nu_{gap}$ were determined by fitting to an asymmetric boxcar function. The change in kinetic inductance was extracted using
\begin{equation}
\label{eq:dLoL}
    \left(\frac{\nu_{gap}(I=0)}{\nu_{gap}(I>0)}\right)^2 = \frac{\Llen(I_{DC})}{\Llen_0}.
\end{equation}
By comparing the measured bandgap frequency at $I_{DC} = 0$ with the circuit model of the transmission line structure, the penetration depth of the NbTiN conductor layer was determined to be 380~nm, and the ratio of the kinetic inductance to the total inductance of the transmission line was estimated to be approximately 0.99.  Neglecting the magnetic inductance, a fit to the data yields
\begin{equation}
\label{eq:nubc}
\frac{\Llen(I_{DC})}{\Llen_0} = 1 + \left( \frac{I_{DC}}{\text{4.3 mA}} \right)^2 + \left( \frac{I_{DC}}{\text{2.8 mA}} \right)^4.
\end{equation} 
The largest current that could be applied before before the transmission line switched to the normal state was 1.5~mA.  At that current, the fractional inductance change $\delta \Llen/\Llen_0$ is 20\%, which is close to the maximum theoretical value.

\begin{figure}
    \centering
    \includegraphics[width=0.47\textwidth]{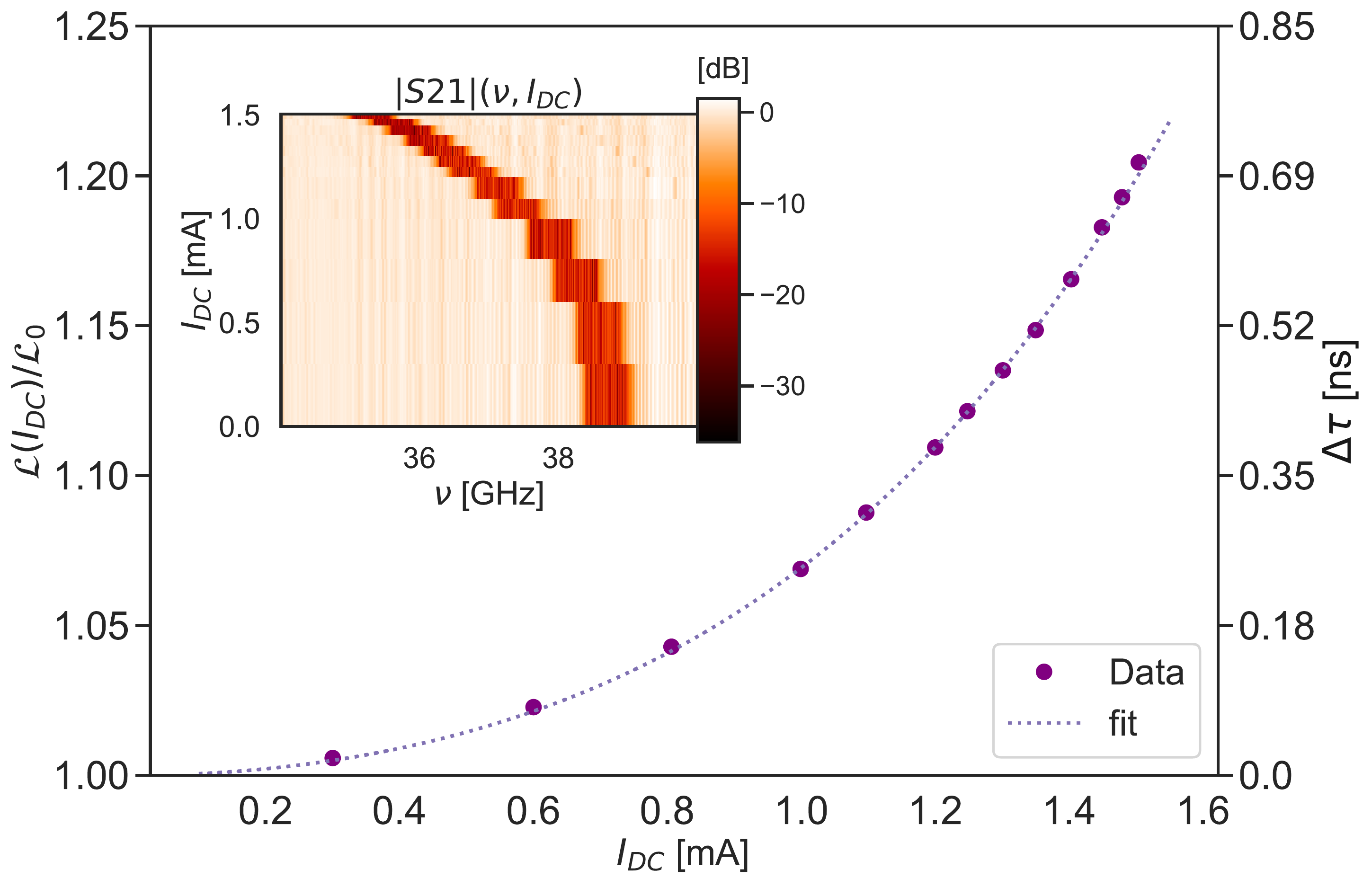}
    \caption{The shift of bandgap is converted as the change of inductance using Eqn.~\ref{eq:dLoL}. The fitting result is shown in Eqn.~\ref{eq:nubc}; this is the left y-axis. The right y-axis is the equivalent current controlled delay $\Delta \tau$ in transmission. The inset shows S21 with respect to $I_{DC}$ and frequency $\nu$, derived from Fig.~\ref{fig:Ka_rawS21}. }
    \label{fig:Ka_bandgap}
\end{figure}

\subsection{Phase delay and on-chip interferometer}

From the designed modulation length and the bandgap frequency measured at $I_{DC}=0$, the zero current propagation velocity is found to be  $v_0=0.0098c$.  The change in the time delay $\Delta\tau = L( \sqrt{\Llen(I_{DC})\Clen} - \sqrt{\Llen_0\Clen})$ through the transmission line reached a maximum of $\sim$0.7~ns at the highest current.  A possible use of this device would be as a controllable delay line in a compact Superconducting On-chip Fourier Transform Spectrometer (SOFTS), operating in the millimeter band~\cite{thakur2020superconducting}. The principle is that a signal is split and fed into two identical current-biased transmission lines. The recombined outputs are terminated on a power detector. When the current is varied in either line, a relative phase-delay is introduced, and the power detector produces an interferogram as function of $I_{DC}$. Microwave circuit design allows for construction of both Mach-Zhender and Michelson interferometers on chip for (sub)millimeter science and other applications.  The frequency resolution will be $1 / \Delta\tau_{max} \sim$1.4~GHz or $0.7$~GHz for single and two-pass circuit designs. Standard optomechanical FTS with GHz resolution are meter-scale, whereas using this concept allows for a millimeter-scale SOFTS~\cite{Pan19}. The physical path length and resolution could be increased while still keeping the area small compared to a pixel in a millimeter-wave imaging array.  Locating at array of SOFTS within the individual feeds of the focal plane an integral field unit could be realized. 

\section{Conclusion}
We have explored the nonlinear kinetic inductance and dissipation of NbTiN microstrip transmission line structures that use a series of microstrip stubs to achieve a 50 Ohm characteristic impedance. Dispersion engineering is realized by changing the length of stubs periodically. At 8.4 GHz, the device loss of 0.35~dB has been measured through a 93mm millimeter length of the microstrip line, corresponding to 333 wavelengths, using an on-chip Fabry–P\'{e}rot interferometer. The low dissipation and large nonlinear current response of these transmission lines make them suitable for realizing traveling wave devices based on nonlinear conversion processes, such as parametric amplifiers.  The low intrinsic dispersion of the superconducting transmission line, and the ability to tune dispersion by varying the geometry, allow for wideband operation of such devices.  As an example, we demonstrate a 7-26~GHz broadband parametric amplifier with gain larger than 15~dB using three-wave mixing.  The magnitude of the current response also makes these transmission lines suitable as current variable delay elements that operate through the millimeter-wave band.

\appendix

\section{Squeezing of quantum fluctuations including loss}
\label{app:asym}

A TWPA with a small amount of loss may function as a quantum limited amplifier as long as the loss can be overcome by the gain of the device and the dissipation of the pump tone does not cause excessive heating. The effect of loss may have more impact on the ability of the TWPA to generate two-mode squeezing. Here we use the results of Ref.~ \cite{Houde:2019paramploss} to estimate the amount of squeezing that might be possible with a microstrip TWPA with the loss that was measured in section \ref{sec:loss}. We assume that the loss is distributed along the length of the TWPA and that the 3WM gain process is phase matched, so Eqn.~\ref{eq:phasematch} is satisfied. To include the asymmetric loss, we define $\alpha_S = \bar{\alpha}-\epsilon$ and $\alpha_I = \bar{\alpha}+\epsilon$, where $\bar{\alpha}$ is the average attenuation factor and $\epsilon$ is the asymmetry. In the low asymmetry limit, the variance of the squeezed quadrature is given by~\cite{Houde:2019paramploss}
\begin{equation}
\label{eqn:Squeeze}
    S \approx  \frac{1}{2(\bar{\alpha}+2 )}\left(\bar{\alpha} + 
                                      \frac{e^{-\bar{\alpha} L }}{2G_{\mathrm{eff}}}\right) 
    +G_{\mathrm{eff}} \frac{\epsilon^{2} e^{-\bar{\alpha} L }}{4 (2 -\bar{\alpha})},
\end{equation}
where 
\begin{equation}
    G_{\mathrm{eff}} = \frac{1}{4}e^{2 L \Delta \phi \sqrt{1+(\epsilon/2)^2}}, 
\end{equation}
and $\Delta \phi=k_p I_{p}I_{DC}/4 I^2_{\dagger}$. $S$ decreases from the zero-point value of 0.5 as the gain increases. For symmetric loss ($\epsilon=0$) the second term in Eqn.~\ref{eqn:Squeeze} is zero and the variance saturates at $S=\bar{\alpha}/2(\bar{\alpha}+2 )$ in the limit of high gain.

\begin{figure}[h!]
    \centering
    \includegraphics[width=0.47\textwidth,trim={0cm 0cm 0cm 0cm},clip]{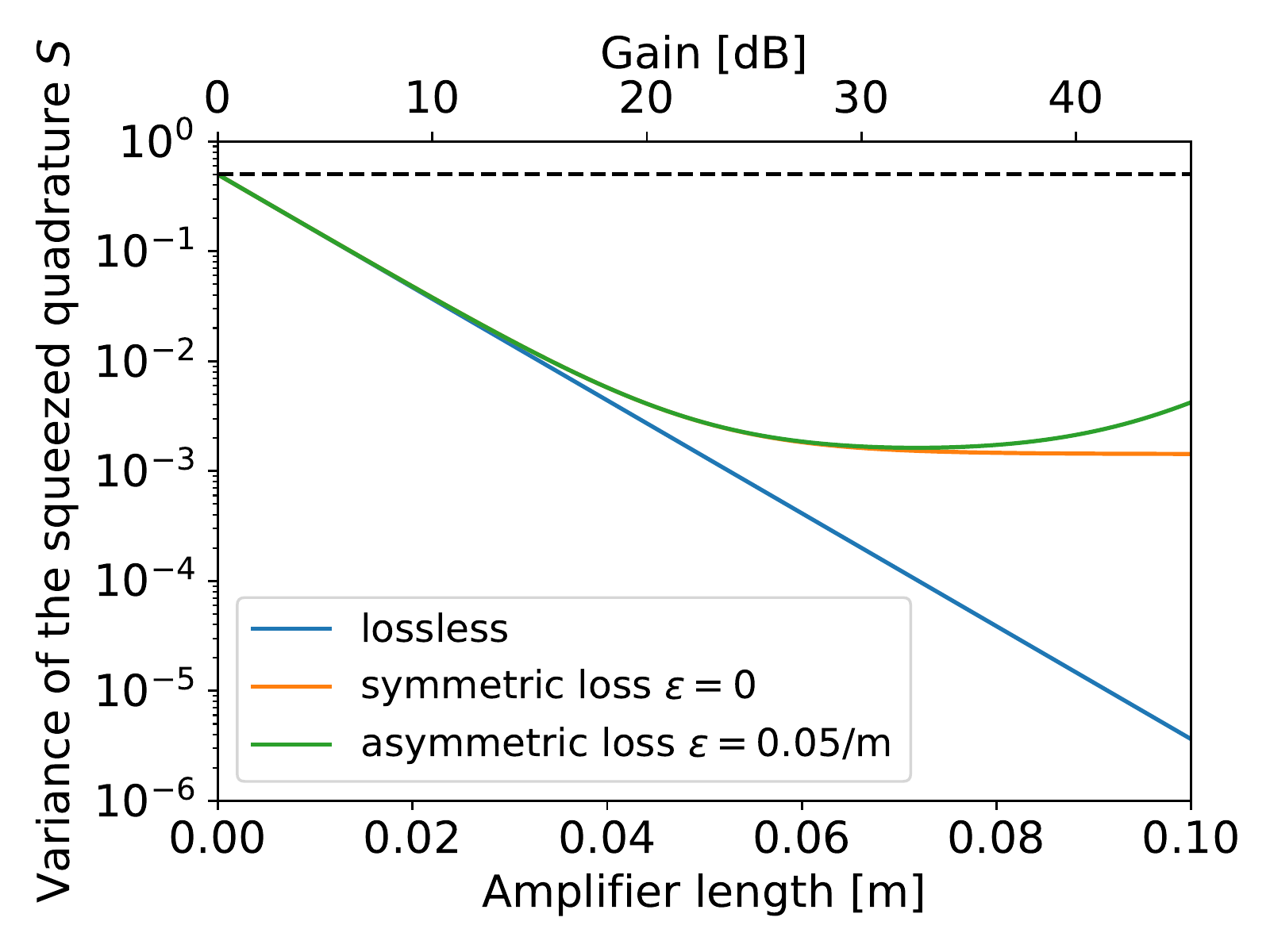}
    \caption{The variance of the squeezed quadrature in three cases: lossless, symmetric loss and asymmetric loss. The dashed line shows $S=0.5$ without squeezing.}
    \label{fig:Sqz_ideal}
\end{figure}

In the case that $Q_i$ is the same for both signal and idler, the loss asymmetry is simply due to the electrical length difference. Using the measured $Q_i=2.8\times10^4$, we have $\bar{\alpha}=0.34/m$, and $\epsilon=0.05$ for a signal at 6~GHz and idler at 8~GHz. We take $I_{DC}/I_*=0.16$ and $I_{p}/I_*=0.037$, similar to the values used in the multi-band gain calculation in Sec.~\ref{sec:gain}. With perfect phase matching, we have $\Delta \phi= 66.85/m$. The variance $S$ is calculated in three cases (Fig.~\ref{fig:Sqz_ideal}). Without loss, the squeezing is proportional to the gain. For symmetric loss, the squeezing saturates with length, as the rate of noise generation balances the squeezing rate. With asymmetric loss, mixing of noise from the amplified quadrature into the squeezed quadrature results in a maximum of the squeezing level at an optimal length~\cite{Houde:2019paramploss}. For the parameters chosen above, the maximum of squeezing is 25~dB at 
\begin{equation}
    L_{\mathrm{opt}} \approx \frac{1}{2\Delta \phi} \log{\frac{\Delta \phi}{2\epsilon }}=54~\mathrm{mm}.
\end{equation}

\section{Generalized coupled mode equations for 3WM and 4WM processes}
\label{sec:coupledmode}

As in the standard treatment of waves interacting in nonlinear optical media, we express
the total current in terms of a number of frequency components,
\be
I = \frac{1}{2} \left( \sum_{n=1}^m A_n(z) e^{i(\beta_n z - \omega_n t)} + {\rm c.c.} \right),
\label{eqn:Isum}
\ee
where the slowly varying complex mode amplitudes $A_n$ satisfy
\be
\left| \frac{d^2 A_n}{dz^2} \right| \ll \left| \beta_n \frac{d A_n}{dz} \right|.
\ee
The number of frequency components included in Eqn.~\ref{eqn:Isum} is chosen to describe the particular
process or processes under investigation. 
For example, a description of parametric amplification in a 3WM medium includes at least three frequencies 
in the sum: the
pump, signal and idler at $\omega_p$, $\omega_s$ and $\omega_i$.  Additional frequencies should be included if they correspond to energetically allowed processes for which one of the input tones has significant power and they are not strongly phase mismatched. In the case of 3WM parametric amplification, second harmonic generation may play a significant role, requiring the inclusion of $2\omega_p$ 
Evolution of the mode amplitudes $A_n$ inside the transmission line
are then found by substituting Eqn.~\ref{eqn:Isum} into \ref{eq:3WM} and matching terms with the same time dependence, resulting
in coupled mode equations for the evolution of the $A_n$. For an arbitrary set of 3WM and 4WM processes connecting $m$ frequency components, these equations can be written compactly as
\begin{equation}
    \frac{dA_n}{dz} = \mathrm{3WM~procs.} + \mathrm{4WM~ procs.},
\end{equation}
where
\begin{equation}
\begin{split}
\mathrm{3WM~procs.} = &\frac{ i \omega_n I_{DC}}{8 I^2_\dagger \bar{c}} 
\sum_{\hspace{-7mm}\begin{array}{c} 
{\scriptstyle 0 \le q_{1\ldots m}, p_{1\ldots m} \le 2 }\\[-7pt]
{\scriptstyle \sum_j p_j + \sum_j q_j  = 2}\\[-0pt]
\end{array}\hspace{-7mm} }
\frac{2!  \, e^{i \beta_n z} }{p_1! \ldots p_m! q_1! \ldots q_m!} \\
& \times \delta(\sum_j p_j \omega_j - \sum_j q_j \omega_j - \omega_n) \\
& \times\prod_{1\le j \le N} \left( A_j e^{-i \beta_j z} \right)^{p_j}
\prod_{1\le j \le N} \left( A_j ^\ast e^{i \beta_j z} \right)^{q_j} ,
\end{split}
\end{equation}
and
\begin{equation}
\begin{split}
\mathrm{4WM~procs.} = &\frac{ i \omega_n}{24 I^2_\dagger \bar{c}} 
\sum_{\hspace{-7mm}\begin{array}{c} 
{\scriptstyle 0 \le q_{1\ldots m}, p_{1\ldots m} \le 3 }\\[-7pt]
{\scriptstyle \sum_j p_j + \sum_j q_j  = 3}\\[-0pt]
\end{array}\hspace{-7mm} }
\frac{3! \, e^{i \beta_n z}}{p_1! \ldots p_m! q_1! \ldots q_m!} \\
& \times \delta(\sum_j p_j \omega_j - \sum_j q_j \omega_j - \omega_n)\\
& \times\prod_{1\le j \le N} \left( A_j e^{-i \beta_j z} \right)^{p_j}
\prod_{1\le j \le N} \left( A_j ^\ast e^{i \beta_j z} \right)^{q_j}.
\end{split}    
\label{eqn:gen4wm}
\end{equation}
The sums in the above equations are over energetically allowed processes and can be inventoried in terms of combinations of integers $p_j$ and $q_j$.  In each equation for $dA_n/dz$, the $p_j$ can be thought of as the numbers of input photons with frequencies $\omega_j$ for a particular process that results in an $\omega_n$ output photon.  The $q_j$ are the numbers of additional output photons at frequencies $\omega_j$ for that process.

\section*{Acknowledgment}

The authors thank Songyuan Zhao and Eduard Driessen for the help in calculations of the Usadel equation.

\input{main.bbl}

\end{document}

%% file: main.bbl
%